\documentclass[twocolumn,showpacs,prl]{revtex4}

\usepackage[dvips]{graphicx}

\begin{document}

\title{Slippery or sticky ! Control of wrinkling patterns by selective adhesion}
\author{Hugues Vandeparre$^{1}$}
\author{Julien L\'eopold\`es$^1$}
\author{Christophe Poulard$^{1}$}
\author{Sylvain Desprez$^{1}$}
\author{Gwennaelle Derue$^{1}$}
\author{Cyprien Gay$^{2}$}
\author{Pascal Damman$^{1}$}
\email{pascal.damman@umh.ac.be}
\affiliation{
$^1$ Laboratoire de Physicochimie des Polym\`eres, Centre d'Innovation et de Recherche en Mat\'eriaux Polym\`eres (CIRMAP), Universit\'e de Mons Hainaut, 20, Place du Parc, B-7000 Mons, Belgium\\
$^2$Centre de recherche Paul-Pascal -- CNRS UPR~8641, Universit\'e de Bordeaux 1, 115 avenue Schweitzer, F-33600 Pessac, France
}

\date{\today}

\begin{abstract}
Wrinkling patterns at the metallized surface of thin polymer films are shown to be sensitive to the sticky or slippery character of the polymer/substrate interface (titanium coating, polystyrene film and coated silicon substrate). Selective prefered wrinkle orientation and amplitude are achieved. Existing theoretical models are expanded to specific boundary conditions (adhesive vs slippery) and rationalize these observations.
\end{abstract}

\pacs{47.54.-r, 46.32.+x, 68.55.-a}
\maketitle

The formation of wrinkles is an ubiquitous phenomenon in Nature\cite{Genzer01}. In most cases, wrinkles appear when a rigid inextensible thin layer, forming the topcoat of an elastic foundation, is subjected to compressive forces. This rather simple concept was adapted to synthetic systems only recently \cite{Bowden01,Shua01,Bowden02,Huck01,Ohzono01,Dalnoki01,Yoo03,Yoo01,Yoo02,Okayasu01,Kwon01}. Although a delicate control of the wrinkling morphogenesis could be extremely useful for many applications in various contrasting domains such as optics\cite{Bowden01,Aschwanden}, microfluidics\cite{Jeon01}, cellular adhesion\cite{Teixeira01}, \ldots 
Thermal compressive stress is one of the easiest way to implement wrinkling but yields disordered patterns (labyrinthal morphology) due to the high symmetry of the stress field. 
Three strategies were proposed in the literature to self-organize the wrinkles: i) engraving the foundation or the substrate with a bas-relief \cite{Bowden01,Bowden02,Ohzono01,Okayasu01}; ii) tailoring the flexural rigidity of the topmost part of the foundation \cite{Huck01,Kwon01}, iii) guiding the wrinkle orientation by pressing a non-planar template against the upper skin during the wrinkle formation \cite{Yoo03}. 

In the present Letter, we will describe an original and easy method to control the spatial layout of wrinkles and thus obtain objects with patterned roughnesses, which could be useful in a number of applications (especially in biophysics areas). To do this, we tune the boundary conditions at the polymer/substrate interface. This is achieved by using chemically patterned substrates with highly contrasted surface free energies ($\gamma$), easily produced by microcontact printing of alkanethiols on gold substrates \cite{Whitesides}. 
Considering the relation between the $\gamma$'s and adhesion, these surfaces are expected to show well-defined slippery and sticky domains. 
At first sight, however, it could seem unrealistic to imagine that the chemical nature of the substrate could affect the elastic instability of the skin through a micron-thick polymer film. This may be the reason why this possibility has not been mentioned in the literature so far. In addition, to explain our original variant of the experiments we will, in this Letter, also expand the existing mechanisms and models described in the litterature to take into account the adhesion at the polymer/substrate interface.

The samples under study consist of a thin layer of high molecular weight polystyrene (PS, $M_n \approx 10^6$ Da) deposited on bare silicon (Si) substrates or chemically patterned gold substrates and further capped with a thin titanium layer (Ti). The PS films were obtained by spin-coating toluene solutions directly onto the different substrates to a thickness ranging from 50 to 1000 nm, as measured by ellipsometry. The Ti layer was deposited onto the polymer surface by thermal evaporation at 0.1 nm/s to a thickness ranging from 10 to 20nm. The bare Si substrates (100) are first cleaned by snowjet and then by ultrasonic treatment in chloroform and methanol for 5 min each. The golden (Au) substrates are prepared by evaporation of a 95 nm thick layer of gold on a Silicon (100) wafer that has been precoated with 5 nm of chromium. These substrates are then cleaned by snowjet and by UV/ozone for 30 min. Finally, heterogeneous samples were created by microcontact printing of octadecanethiol chemical lines. For both systems, wrinkling is induced by heating the bilayer samples at 130$^\circ$C, i.e., 30$^\circ$C higher than the glass transition temperature, $T_g$, of the polymer layer for a short period of time, {\em ca.} 1~min. This temperature is reached at a very high rate (120$^\circ$C/min). The resulting metal surface structures were examined by optical microscopy (OM) and Atomic Force Microscopy (AFM). 

Metal/polymer/substrate trilayers have been reported~\cite{Yoo01,Yoo02,Okayasu01} to exhibit wrinkling when heated above the $T_g$ of the polymer (e.g., polystyrene). Concerning the origin of wrinkling, despite the general agreement on the importance of compressive stresses arising from the mismatch in expansion coefficients for the different layers, there is some debate~\cite{Yoo01,Yoo02,Okayasu01} on the exact mechanism.
In the following, we present an original variant of the experiment and show in detail that the mechanism described by Okayasu {\em et al.}~\cite{Okayasu01} and some aspects of the model by Cerda and Mahadevan~\cite{Mahadevan} can be expanded to account for the observed effect of boundary conditions at the polymer/substrate interface.

\begin{figure}
\centering \includegraphics[width= 4.5cm]{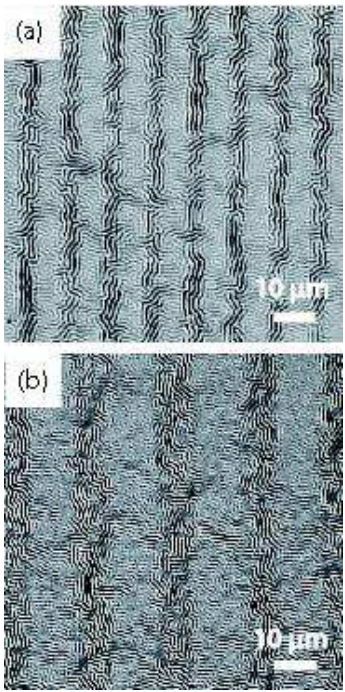}
\caption{Optical Micrographs of wrinkled surfaces obtained with a 20~nm thick metal layer and a 60~nm polymer layer on chemically patterned gold substrates, after annealing at 130$^{\circ}$C . Alternating stripe widths are (a) 5$\mu$m/5$\mu$m, (b) 10$\mu$m/10$\mu$m.}
\label{pattern}
\end{figure}

In contrast to previous studies made on bare silicon, we focus on substrates with alternating areas of high and low adhesion with PS. Those areas are stripes (of width $\Lambda$) of either bare gold (adhesive regions) or octadecanethiol treated gold (ODT, slippery regions). 
Surprisingly, the kinetics of wrinkle growth differs on both stripes. Even though the temperature ramp imposed to the sample is rather fast (120$^{\circ}$C/min), the wrinkles develop first above slippery regions (around 125$^{\circ}$C), and somewhat later above adhesive regions (around 130$^{\circ}$C). 
As shown in Figure~\ref{pattern}, the wrinkle morphology and orientation are also markedly different between sticky and slippery regions: i) observed patterns combine the substrate periodicity ($\Lambda$) and wrinkles ''natural'' wavelength ($\lambda$), ii) the wrinkles are preferentially aligned parallel to the chemical stripes on slippery regions (ODT) and perpendicular to the stripes on adhesive regions (Au), and iii) the wrinkle amplitude $A$ is smaller above adhesive (Au) regions ({\em ca.} 15~nm) than above slippery (ODT) regions ({\em ca.} 25~nm).

Note that such unambiguous experimental results cannot be explained by the mechanisms and models proposed recently \cite{Okayasu01,Yoo01,Mahadevan} and highlight the need for a more thorough understanding of wrinkling. Before discussing the formation of wrinkles on heterogeneous surfaces, we will first discuss what could be the influence of adhesion ({\em i.e.,} sticky {\em vs.} slippery substrates) on the growth of wrinkles for {\em homogeneous} substrates.

For such trilayer systems, the origin of the wrinkles is usually related to the mismatch in thermal expansion coefficients between the polymer foundation and the thin metal layer. But this, by itself, does not constitute an explanation for the wrinkling phenomenon, as pointed out by Yoo and Lee~\cite{Yoo01}. Indeed, the (linear) expansion coefficient of PS ($\alpha_{PS} \sim 5\,10^{-4} K^{-1}$ above $T_g$) is much larger than that of Ti ($\alpha_{Ti} \sim 8\,10^{-6} K^{-1}$). Hence, we could na\"ively consider that upon heating the system above the PS glass transition temperature $T_g$, the Ti membrane should be set under {\em tension} rather than under compression.

In fact, as suggested by Okayasu {\em et al.}\cite{Okayasu01}, there is also a thermal expansion mismatch between the Ti membrane and the Si substrate ($\alpha_{Si} \sim 3\,10^{-6} K^{-1}$), and this mismatch, by contrast with the Ti/PS mismatch, favours the {\em compression} of the Ti membrane. Paradoxically, although the Ti/Si mismatch is much weaker than the Ti/PS mismatch, we will show that it is dominant and causes the Ti membrane to buckle.

\begin{figure}
\centering \includegraphics[width= 7cm]{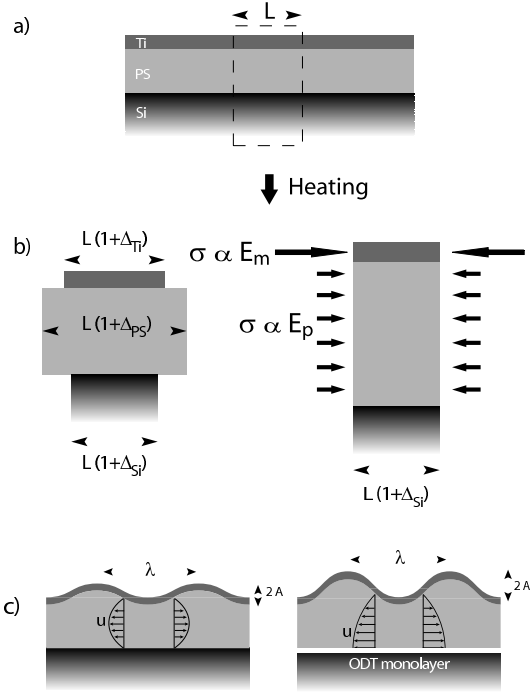}
\caption {Schematic representation of wrinkle formation on a bare substrate. a) Layered system prior to heating. b) Upon heating, (left) expansion that would be obtained for each layer independently in the absence of any surrounding material and (right) distribution of the compressive stresses in the trilayer prior to wrinkling. c) Deformation profile in the polymer layer induced by wrinkling in adhesive (left) and slippery regions (right).}
\label{scheme}
\end{figure}

Without any surrounding material, each layer, depicted on Figure~\ref{scheme}a, would expand isotropically by a factor $\Delta_i(T) =\int_{T_0}^{T} \alpha_{i}\;dT$ (Figure~\ref{scheme}b, left), where $T_0$ is the reference temperature, at which the system is stress-free, namely room temperature (if we neglect internal stresses due to solvent evaporation), $i$ representing the metal, the polymer and the substrate.
The elastic moduli of the Ti membrane and of the Si substrate are obviously much larger than that of the PS layer, especially above $T_g$: $E_s\sim E_m\gg E_p$. The thicknesses also differ substantially: $H_s\gg H_p > h_m$. As a result, the Si substrate is by far the most prominent layer in the thermal expansion process: not only will the overall linear (in-plane) expansion of the trilayer system almost exactly match the spontaneous value for the substrate ($\Delta_s$), but the high flexural substrate rigidity (proportional to $E_s H_s^3$) will hinder any attempt to relieve stresses through bending of the trilayer as a whole. As a result, the metal membrane is compressed (expansion $1-\Delta_m+\Delta_s < 1$) with in-plane compressive stress $2E_m (\Delta_m - \Delta_s) = 2E_m \delta$ (while there is no stress in the out-of-plane direction) and elastic energy $U_c \sim 2E_m h_m \delta^2$.

\begin{figure}
\centering \includegraphics[width= 7cm]{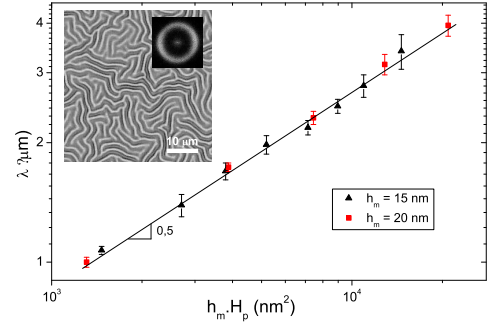}
\caption{Evolution of the wavelength versus the product of the metal and polymer thickness for trilayers Ti/PS/Si heated at 130 $^{\circ}$C. The inset shows OM image of a wrinkled surface obtained with a 15~nm thick metal layer and a 350~nm thick polymer layer (the FFT of a broad sample area, 10 times larger than the OM image is also shown).}
\label{isotropic_wrinkles}
\end{figure}

The random wrinkling pattern observed upon heating our samples has a well-defined characteristic wavelength (see inset in Figure~\ref{isotropic_wrinkles}) which is much larger than the PS film thickness ($\lambda\gg H_p$). To determine the criterion for the selection of the wrinkles wavelength, we expand the scaling energetic approach proposed by Cerda and Mahadevan~\cite{Mahadevan}. In the presence of wrinkling, thanks to the corresponding increase in curvilinear length, the membrane compression is partly relieved. It is reduced by $\pi^2 A^2 / \lambda^2$ for a sinusoidal membrane deformation with amplitude $2A$. This obviously lowers the compression elastic energy :
\begin{equation}
U_c \sim 2 E_m h_m (\delta-\pi^2 A^2/\lambda^2)^2<  2E_m h_m \delta^2 
\label{Uc}
\end{equation}

Additionally, the energy of the system also includes the membrane bending energy $U_b$ and the polymer elastic energy $U_p$ (the foundation must follow the sinusoidal deformation of the membrane). 

The bending energy of the Ti layer, expressed per unit surface, is determined~\cite{Mahadevan} by the membrane curvature (typically $A/\lambda^2$):
\begin{equation}
\label{Ub}
U_b \sim E_m h_m^3 \left(\frac{A}{\lambda^2}\right)^2
\end{equation}
where $E_m h_m^3$ is the order of magnitude of the membrane bending rigidity.

As we shall now see, the elastic energy stored in the polymer foundation due to the sinusoidal membrane deformation depends on boundary conditions. Let us consider two limiting cases: no-slippage (sticky) and full-slippage (slippery), assuming that the true mechanical behavior at the polymer/substrate interface will be somewhere in between (more sticky on bare Au and more slippery on ODT). Because the polymer is incompressible, the membrane vertical displacement induces horizontal displacements in the polymer layer. Although the maximum horizontal displacement in the polymer layer is independent of boundary conditions for a given amplitude and wavelength ($u \sim A\lambda/H_p$), the displacement profiles differ, as sketched on Fig.~\ref{scheme}c: the maximum displacement is located at mid-height in the case of a sticky interface, while it is located at the polymer/substrate interface if it is slippery. 
As a result, the shear deformations $\partial u/\partial z$ (which are on the order of $u/H_p \sim A\lambda/H_p^2$) are on average twice as large in the sticky case than in the slippery case. Correspondingly, the elastic energy per unit surface area $U_p \sim E_p \int (\partial u/\partial z)^2\,{\rm d}z$ is larger by a factor 4 in the sticky case. In the following expression for the polymer elastic energy, we therefore include a numerical factor $K$, where it is understood that $K^{stick}/K^{slip}=4$:
\begin{equation}
\label{Up}
U_p = K E_p H_p \left(\frac{A\lambda}{H_p^2}\right)^2 
\end{equation}
Note that in the opposite case of a thick polymer film ($\lambda\ll H_p$), $H_p$ must be replaced by $\lambda$ in the above expression~\cite{Mahadevan} and the boundary condition at the lower interface has no impact.

Minimization of the total energy $U_c+U_b+U_p$ yields the wavelength, and the critical deformation, $\delta_c$:
\begin{eqnarray}
\label{lambda}
\lambda &\sim& \left(\frac{1}{K}\right)^{1/6} (h_m H_p)^{1/2} \left(\frac{E_m}{E_p}\right)^{1/6} \\
\label{delta}
\delta_c &\sim& K^{1/3} \frac{h_m}{H_p} \left(\frac{E_p}{E_m}\right)^{1/3}
\end{eqnarray}

As shown in Figure~\ref{isotropic_wrinkles}, the observed dependence of the wrinkle wavelength on the metal and polymer thicknesses is adequately described by the allometric relation $\lambda \propto (h_m H_p)^{1/2}$ given by (Eq.~\ref{lambda}).

Let us now return to heterogeneous substrates with slippery and sticky stripes. The most striking result of the model is probably the influence of the boundary conditions on the critical stress. Indeed, as $\delta_c$ is proportional to $K^{1/3}$ (see Eq.~\ref{delta}), the threshold compression for wrinkling is expected to be lower for slippery substrates (such as ODT) than for adhesive ones (such as bare Au).

\begin{figure}
\centering \includegraphics[width=7cm]{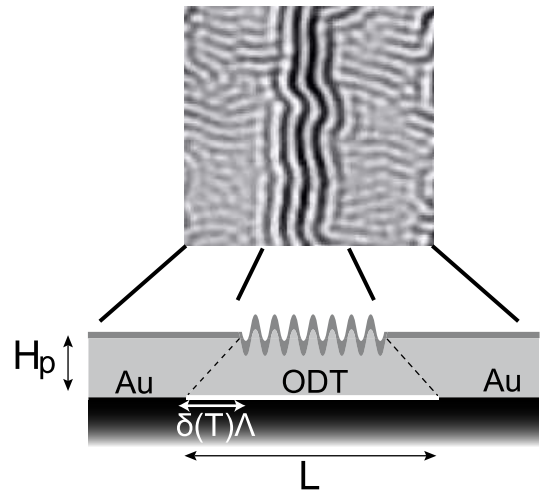}
\caption{Schematic representation of the slight membrane displacement (and corresponding deformation of the polymer layer) when wrinkles appear above slippery regions. This displacement induces the observed prefered wrinkle orientation above slippery stripes (and later the observed opposite prefered orientation above sticky stripes). It is also at the origin of the contrasted wrinkle amplitude between both types of regions.}
\label{profile}
\end{figure}

The fact that wrinkles appear first in the slippery regions will now help us to understand: {\em i)} the prefered wrinkle orientation and {\em ii)} the difference in wrinkle amplitudes in both regions.
Before wrinkles appear (say, at 120$^{\circ}$C), the in-plane compressive stress in the membrane is isotropic. Once wrinkles appear in the slippery regions (at {\em ca.} 125$^{\circ}$C), the compressive stress is partly relaxed there. The (yet undeformed) membrane in the {\em sticky} regions now tends to expand laterally at the expense of the wrinkled membrane in the slippery regions. Hence, at the border between slippery and sticky stripes, the membrane is slightly displaced towards the slippery regions (see Fig.~\ref{profile}).
In the case of rather narrow stripes, this displacement is expected to be on the order of $[\Delta_m(T)-\Delta_s(T)]\Lambda=\delta(T)\Lambda$. This slight displacement favours wrinkles oriented parallel to the stripes above the slippery regions, while the membrane diplacement in the direction of the stripes remains unnoticable.
By contrast, when wrinkles later appear in the sticky regions (at {\em ca.} 130$^{\circ}$C), the membrane displacement at the stripe border causes the compressive stress to be weaker in the direction perpendicular to the stripes than in the parallel direction. The new wrinkles thus tend to develop perpendicular to the stripe direction.
In addition, the slight membrane displacement at the stripe border induces a stronger compression in the slippery regions, resulting in a larger wrinkle amplitude~\cite{Mahadevan}, as indeed observed, see Fig.~\ref{pattern}.

%
We have explained in detail what could be the origin of the in-plane compressive stress that induces wrinkling in thin metal/polymer bilayers deposited on a thick substrate. We also showed a new and simple approach to {\em i)} orient wrinkling patterns in thin polymer/metal bilayers by chemically patterning the {\em substrate} with regions of high and low adhesion and {\em ii)} elaborating patterns with contrasted roughnesses. In this way, a mere change in the surface tension of both slippery and sticky regions makes it possible to obtain a well-controled roughness ratio between both areas. Such controled roughness patterns would be useful in a number of areas, including biophysics. Existing models were expanded to rationalize the observed impact of boundary conditions at the polymer/substrate interface on surface 
wrinkling.

P.D. and H.V. thank Dr. A.~Boudaoud, B.~Audoly and R.~Lazzaroni for stimulating discussions. The authors thank J. Piotte for his contribution. The Belgian National Fund For Scientific Research (FNRS), the FRIA and the Walloon region (research project CORRONET) are acknowledged for financial support. P.D. is a Research Associate of the FNRS. C.G. gratefully acknowledges support from  the FNRS and hospitality from LPCP.

\end{document}